\documentclass[11pt]{article}
\usepackage{amsmath}
\usepackage{amssymb}
\usepackage{graphicx}
\usepackage{multicol}
\usepackage{indentfirst}
\usepackage{caption}
\usepackage{placeins}
\usepackage{subcaption}
\usepackage{float} 

\usepackage{calc}
\usepackage[margin=0.5in]{geometry}

\usepackage{color}
\usepackage[usenames]{xcolor}

\usepackage[cp1252]{inputenc}
\usepackage{listings}

\begin{document}
\title{Ultrafast Charge Transfer Enhancement in CdS-MoS$_2$ via Linker Molecule}
\author{Matthew Ciesler, Han Wang, Shengbai Zhang, and Damien West}
\maketitle
\vspace{-0.75cm}{\small Department of Physics, Applied Physics and Astronomy, Rensselaer Polytechnic Institute, Troy, NY 12180}

\begin{abstract}\noindent
Hybrid systems, which take advantage of low material dimensionality, have great potential for designing nanoscale devices. Quantum dots (QDs)---a 0D nanostructure---can be combined with 2D monolayers to achieve success in photovoltaics and photocatalytic water splitting. In such colloidal systems, ligand molecules such as cysteine play an important role in device performance. The role of the ligand molecule in these QD heterostructures is poorly understood. In this study, time-dependent density functional theory (TD-DFT) is employed in order to explore how the ligand affect the charge transfer at the ultra-fast timescale. We study the charge transfer dynamics in CdS-MoS$_2$ heterostructures both with and without an organic linker molecule. We find that the ligand molecule enhances the ultrafast charge transfer, and that electrons are preferentially transferred from CdS to MoS$_2$ as band alignment would predict. The electronic dynamics and time-evolved projection character are sensitive to the ionic temperature and excitation density.
\end{abstract}
\begin{multicols}{2}

\section{Introduction}
Within the emerging solutions for alternative energy sources, photovoltaics \cite{kamat2013quantum}\cite{guijarro2010direct}\cite{kamat2012boosting} and photocatalyzed water splitting \cite{ar2020cdse}\cite{luo2020hybrid}\cite{benck2014catalyzing} based on quantum dots (QDs) have great potential. In these third generation devices, the operational principle is that the QD absorbs light, creating photoexcited electrons and holes which differentiate to the electrode and counter-electrode respectively. This separation of charge is fundamental to the operation of the device, and so understanding the nature of this charge separation is critical to the understanding of these systems.

One important facet of the electron transfer process in these QD systems is the ultrafast dynamics \cite{huang2008photoinduced}\cite{huang2010multiple}\cite{yang2012strong}\cite{kaniyankandy2012ultrafast}. At this timescale, the mechanism of photoexcited charge separation is important, together with the interaction with the QD surface and charge trapping states \cite{harvie2018ultrafast} \cite{goodman2018ultrafast}. In the case of bifunctional ligand molecules\cite{talapin2010prospects}, the ligand is expected to play a role in enhancing the charge transfer. In this study, we focus on the ligand effects in a CdS/MoS$_2$ heterostructure, which (along with closely related systems such as CdSe/MoS$_2$) have seen success in photovoltaics, photocatalysis, and photodetection \cite{frame2010cdse}\cite{yuan2020growth}\cite{wang2019synthesis}\cite{dong2015efficient}.
is a window into a relatively unexplored regime for for the colloidal QD system, because TD-DFT
Time-dependent density functional theory (TD-DFT) can allow one to simulate these nanoscale systems at a timescale of femptoseconds, which can capture important charge transfer dynamics \cite{benko2002photoinduced}. TD-DFT is well-suited to study dynamic processes such as excitations \cite{castro2006octopus}, and real time integration schemes \cite{huix2020time} are especially suited for studying photoexitations \cite{lian2018photoexcitation}. From a theoretical standpoint, the Runge-Gross theorem \cite{runge1984density} demonstrates that TD-DFT is an \emph{ab initio} technique that, in principle, is as fundamental as the Shr\"odinger equation

\section{Methods}

For time-dependent DFT (TD-DFT) calculations, a modified version of the Spanish Initiative for Electronic Simulations with Thousands of Atoms (SIESTA) \cite{soler2002siesta} software package was employed. The basis for this calculation is a linear combination of atomic orbitals (LCAO), consisting of double zeta plus polarization (DZP) orbitals. In calculations at a nonzero temperature, molecular dynamics (MD) are performed in an NVE ensemble (microcanonical ensemble), where the system is thermalized by giving the ground state atomic configuration an appropriate amount of kinetic energy, and then evolving the system over time. The timesteps used for dynamic simulations are 1 fs for MD, and 0.5 Rydberg time (0.024 fs) for TD-DFT.

Regarding the atomic structure (See Fig. \ref{fig:tddft-structs}) of the simulation, note that the thiol group of ligand molecules such as cysteine coordinates to surface Cd \cite{watson2010linker} \cite{chung2008structures} \cite{kurihara2019capping} in CdSe/CdS QDs. Due to this preferrential binding, the Cd-terminated surface \cite{rempel2006density} plays an especially important role. On the other side of the heterostructure, the pristine MoS$_2$ surface is inert and is functionalized with a S vacancy. These S vacancies are relatively abundant in real MoS$_2$ samples \cite{qiu2013hopping} \cite{liu2013sulfur}, and functionalization to this defect via an O atom has been previously reported \cite{kim2014tunable}. Thus, the ligand will be bifunctional to CdS and MoS$_2$ via its S an O ends, respectively. This three-way heterostucture construction is similar to previous DFT simulations \cite{ciesler2021ligand}. For lattice commensuration, we used a $4\times4$ cross-section of CdS combined with a $2\sqrt{7}\times2\sqrt{7}$ cross-section of MoS$_2$, and strained MoS$_2$ by 2.8\%.

\section{Results}

\begin{figure}[H]
\begin{center}
\begin{subfigure}[b]{0.23\textwidth}
\includegraphics[width=\textwidth]{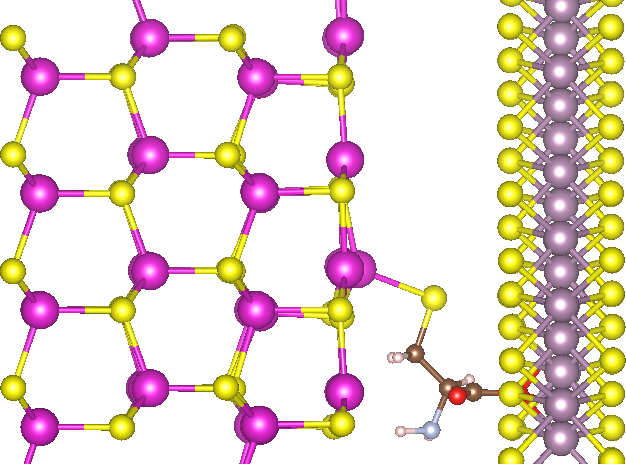}
\vspace{-25pt}\caption{}\vspace{7pt}
\includegraphics[width=\textwidth]{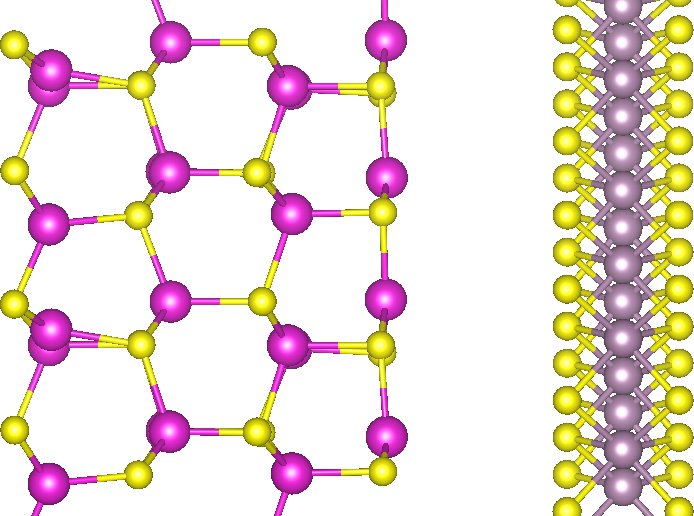}
\vspace{-25pt}\caption{}\vspace{7pt}
\end{subfigure}
\hspace{0.01\textwidth}
\begin{subfigure}[b]{0.23\textwidth}
\includegraphics[width=\textwidth]{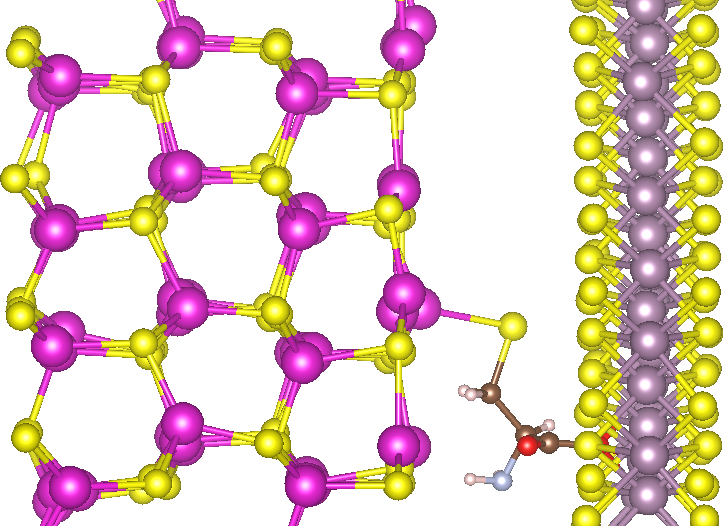}
\vspace{-25pt}\caption{}\vspace{7pt}
\includegraphics[width=\textwidth]{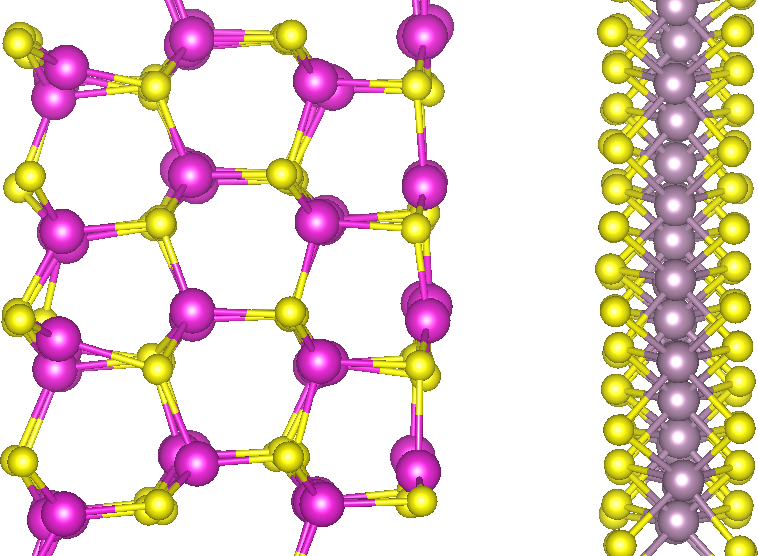}
\vspace{-25pt}\caption{}\vspace{7pt}
\end{subfigure}
\end{center}
\caption{Atomic structure for the heterostructures used in TDDFT. The initial ionic temperature was (a-b) 0K and (c-d) 300K. The heterostructures are (a,c) CdS-Cys($r$)-MoS$_2$-v$_S$ and (b,d) CdS--MoS$_2$. Atoms: H (white), C (brown), O (red), N (blue), Cd (purple), S (yellow), Mo (grey).}
\label{fig:tddft-structs}
\end{figure}

\begin{figure}[H]
\begin{center}
\begin{subfigure}[b]{0.23\textwidth}
\includegraphics[width=\textwidth]{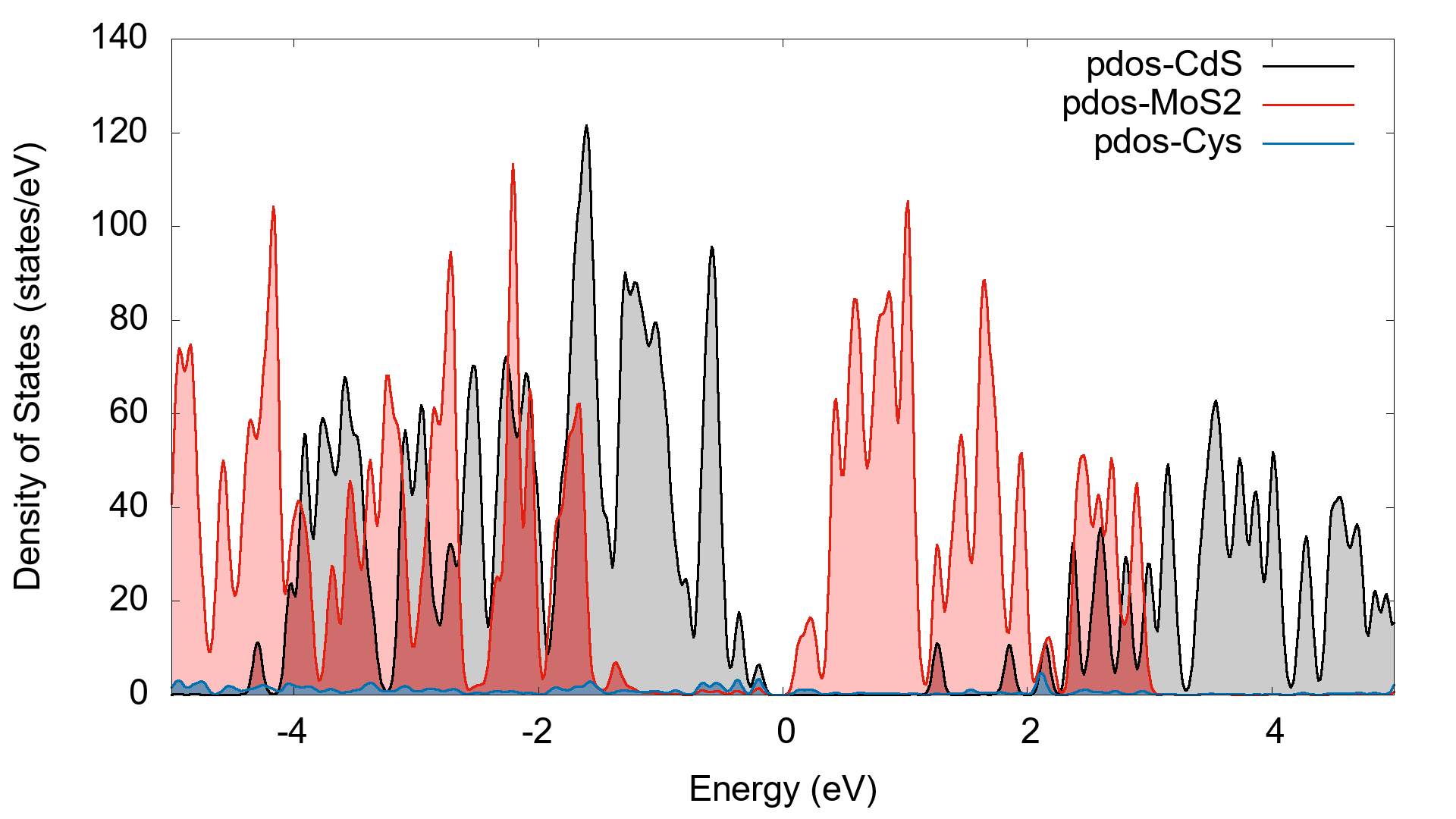}
\vspace{-35pt}\caption{}\vspace{15pt}
\includegraphics[width=\textwidth]{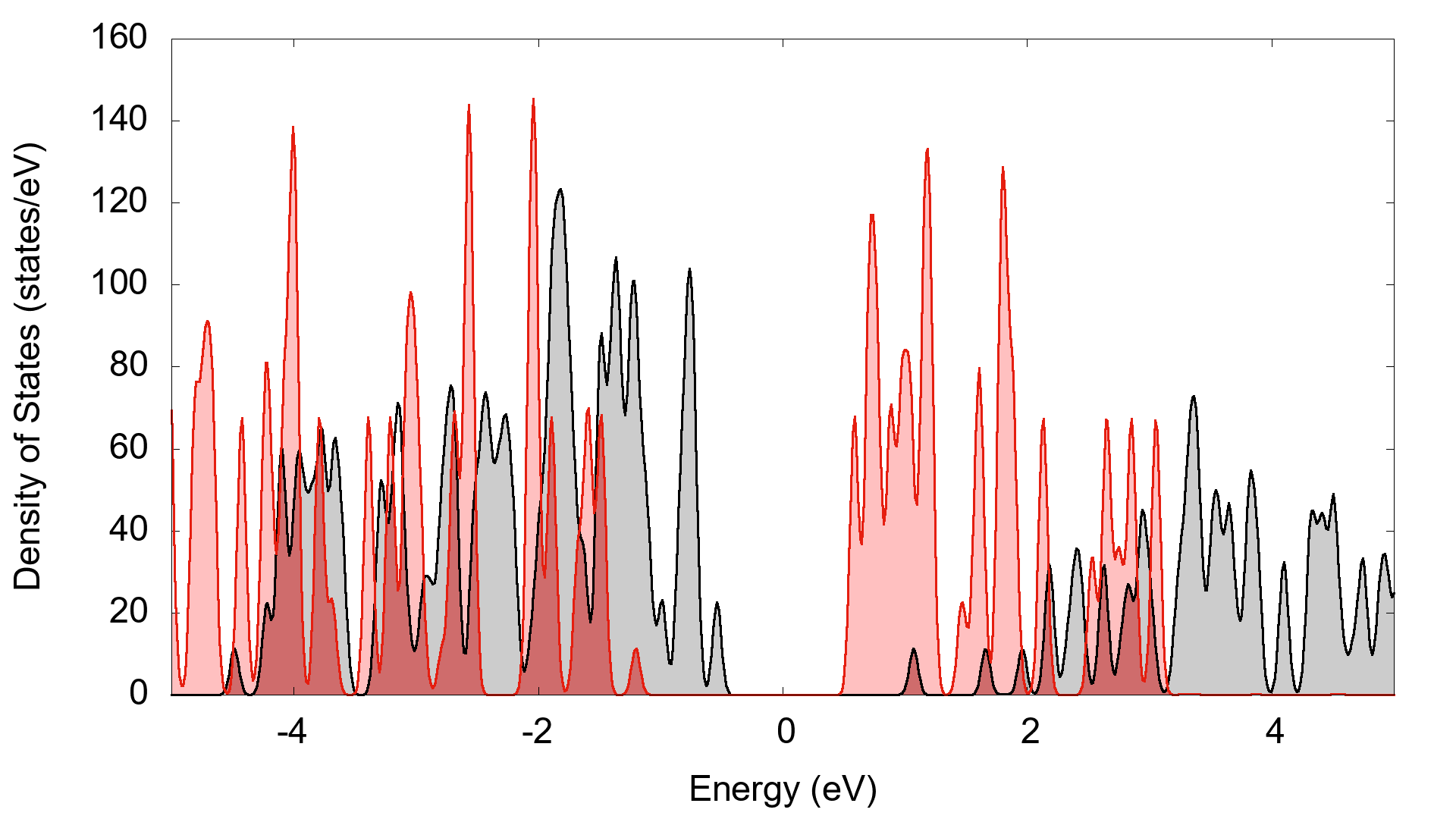}
\vspace{-35pt}\caption{}\vspace{15pt}
\end{subfigure}
\hspace{0.01\textwidth}
\begin{subfigure}[b]{0.23\textwidth}
\includegraphics[width=\textwidth]{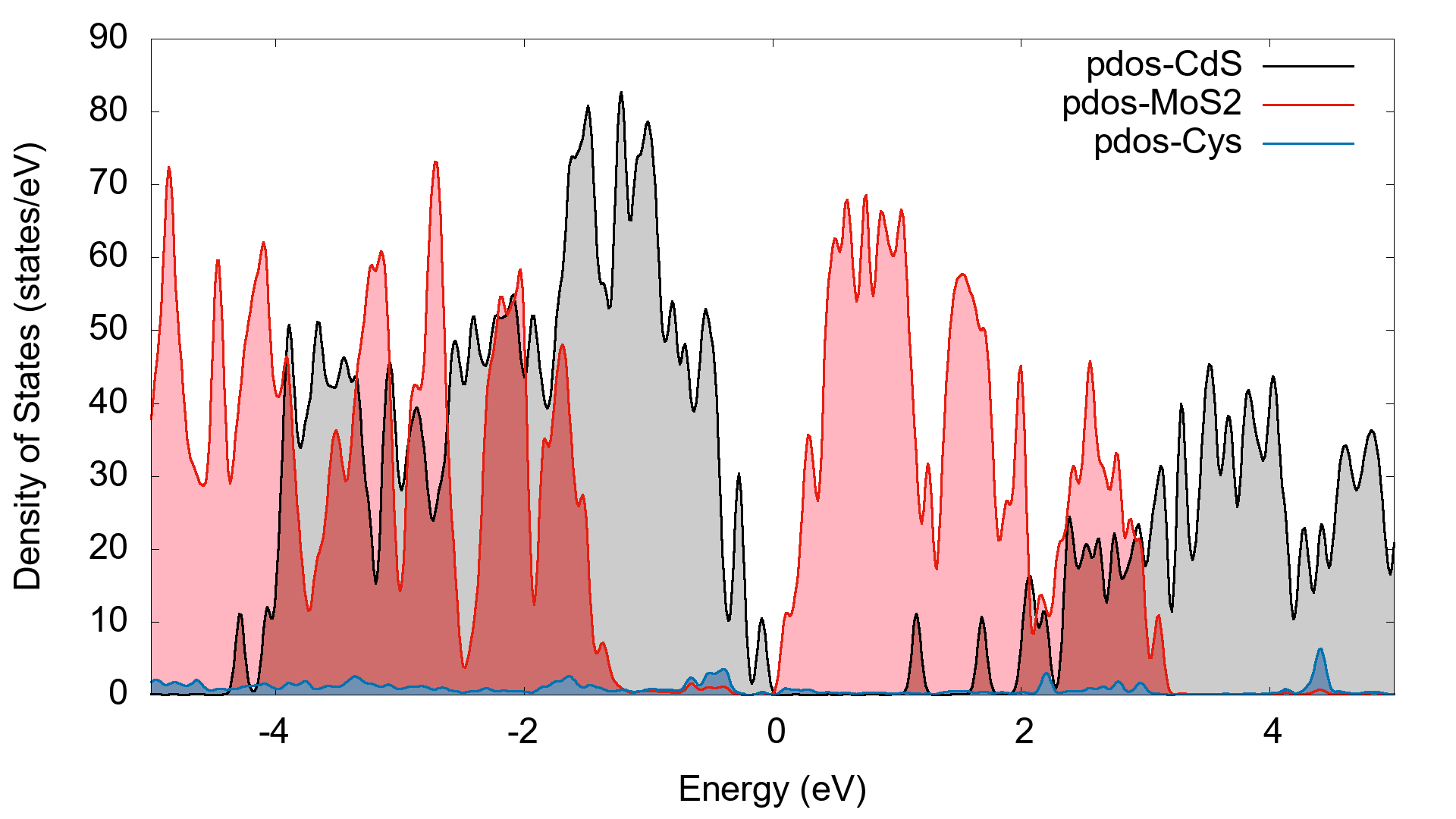}
\vspace{-35pt}\caption{}\vspace{15pt}
\includegraphics[width=\textwidth]{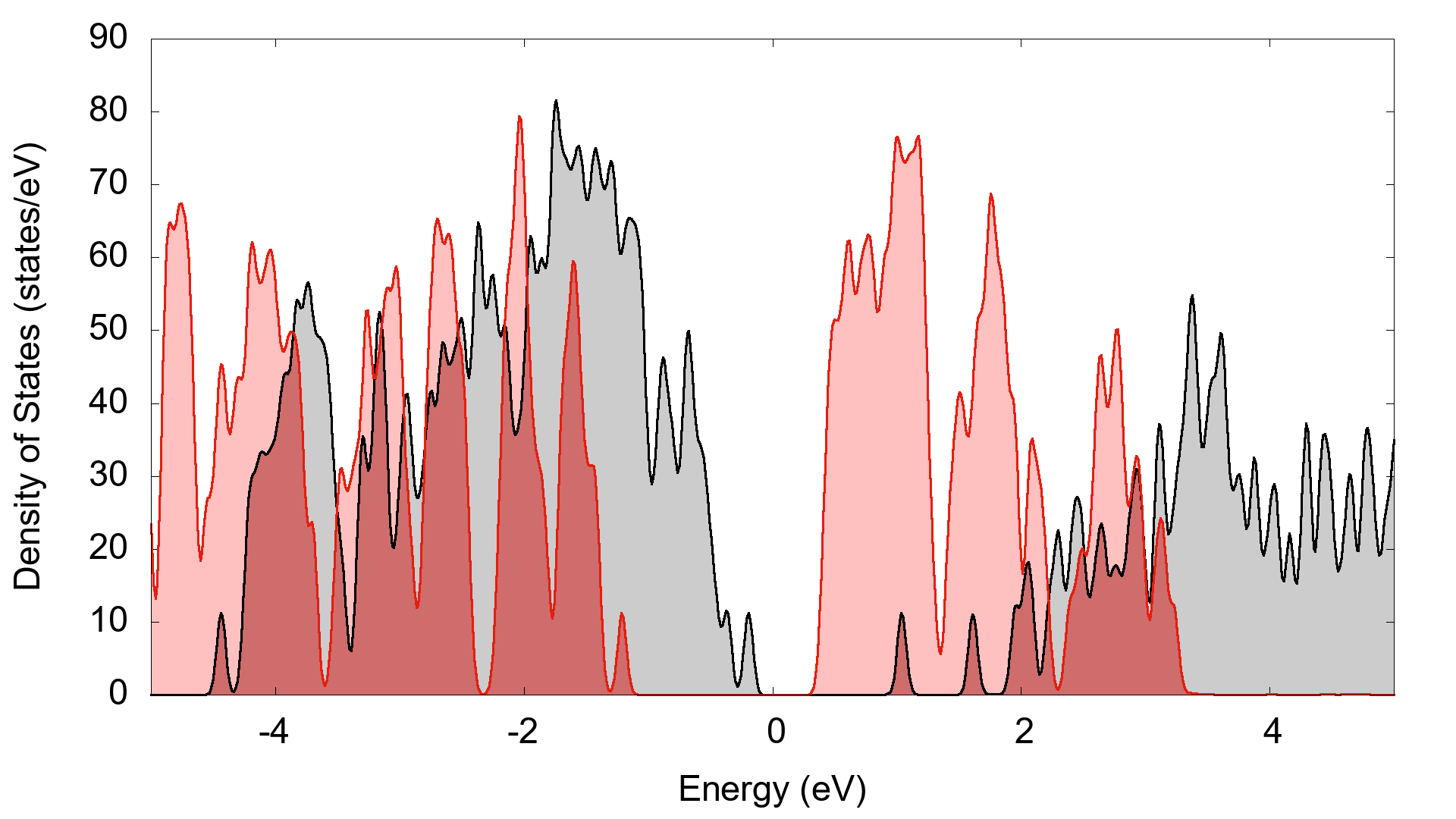}
\vspace{-35pt}\caption{}\vspace{15pt}
\end{subfigure}
\end{center}
\caption{Density of states for the heterostructures used in TDDFT. The initial ionic temperature was (a-b) 0K and (c-d) 300K. The heterostructures are (a,c) CdS-Cys($r$)-MoS$_2$-v$_S$ and (b,d) CdS--MoS$_2$.}
\label{fig:tddft-pdos}
\end{figure}

Fig. \ref{fig:tddft-structs} shows the atomic structures for the heterostructures used in the TD-DFT simulations. The 0K (300K) separated systems are constructed so that the CdS and MoS$_2$ films are relaxed (thermalized) separately and then combined afterwards, respectively. The density of states is also plotted in Fig.\ref{fig:tddft-pdos}. In all of the cases, as can be seen in the density of states, the doubly-coordinated cysteine ligand has its gap opened up by the interactions, and type-II alignment occurs between CdS and MoS$_2$, although for case (c) the gap between the CdS VBM and the MoS$_2$ CBM is very small. The DOS for the 300K systems are broadened compared to the 0K systems, and in the systems without a ligand molecule, the change in the dipole causes the alignment between CdS and MoS$_2$ to shift so that the HOMO-LUMO gap of the system increases.

\subsection{Results for 300 K Cys Heterostructures}

Beginning with the 300K results, Table \ref{table:tddft-proj} shows the process for selecting the excitation states. One band is chosen as the excited electron state, which has to be empty initially, and another state for the hole, which is initially full. The criteria for the state to use is that the state should first and foremost be mostly projected onto CdS. This is because the excitation that the simulation is trying to mimic is a photoexcitation in CdS. But, as we see from the table, no state is purely CdS and there will always be a small amount of delocalization to Cys and MoS$_2$.

\begin{table}[H]
\centering
\begin{tabular}{|c|c c c c c|} 
\hline
Band & CdS & Cys & MoS$_2$ & E (eV) & ID \\
\hline
806 & 98.8 & 1.0 & 0.2 & -0.336  & CdS VBM-3 \\
807 & 99.5 & 0.4 & 0.1 & -0.235 & CdS VBM-2 \\
809 & 88.7 & 6.0 & 5.2 & -0.075 & HOMO \\
810 & 2.4 & 8.9 & 88.7 & 0.056 & LUMO \\
901 & 99.8 & 0.1 & 0.2 & 2.156 & CdS CBM+2 \\
902 & 99.2 & 0.3 & 0.5 & 2.207 & CdS CBM+3 \\
\hline
\end{tabular}
\caption{Projection character (CdS, Cys, MoS$_2$, in percent), energy (E), and identity of states in the $T=300K$ CdS--Cys($r$)--MoS$_2$-v$_S$ heterostructure.}
\label{table:tddft-proj}
\end{table}

\begin{figure}[H]
\begin{center}
\begin{subfigure}[b]{0.306\textwidth}
\includegraphics[width=\textwidth]{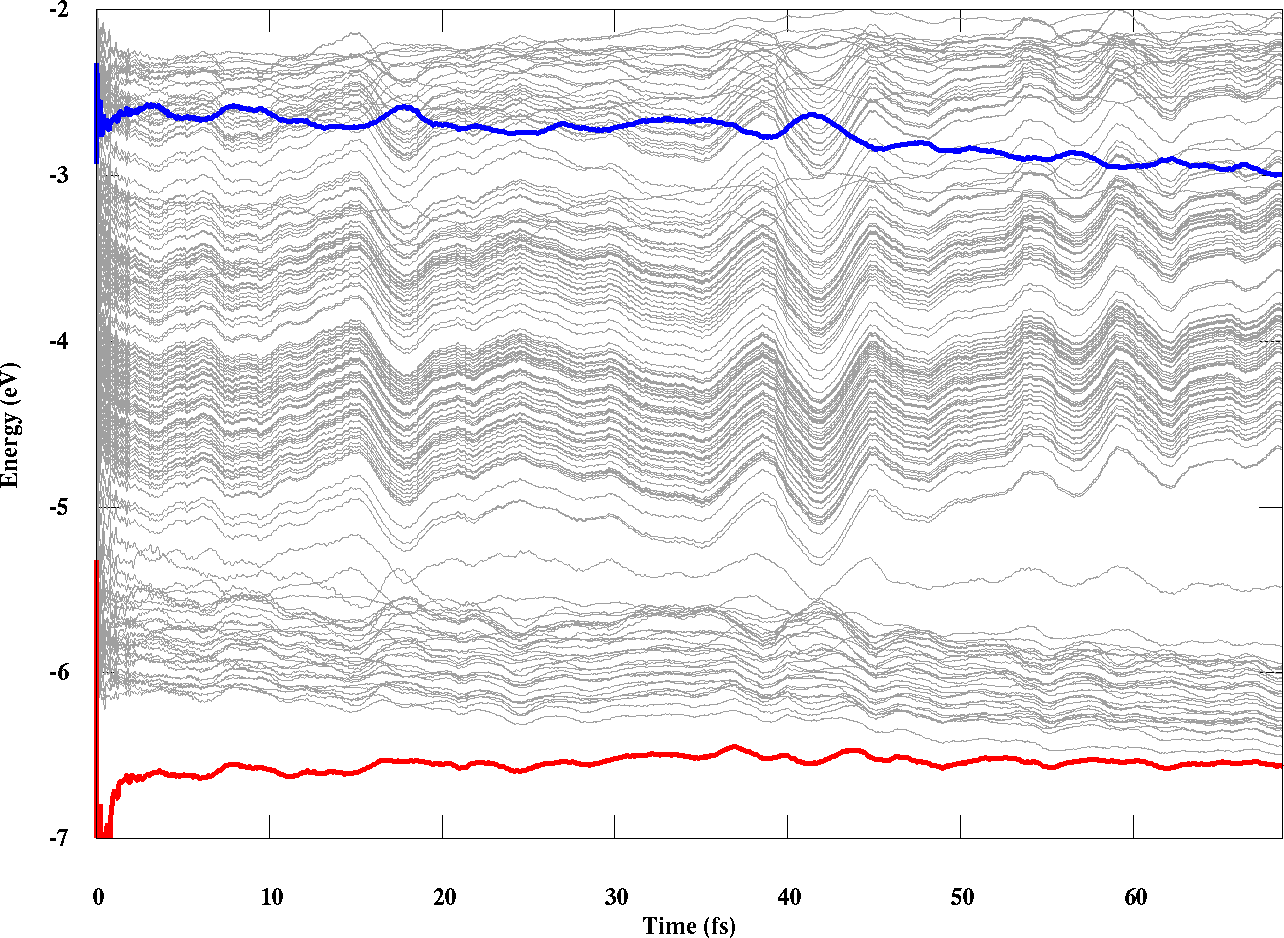}
\vspace{-60pt}\caption{}\vspace{23pt}
\end{subfigure}
\begin{subfigure}[b]{0.162\textwidth}
\includegraphics[width=\textwidth]{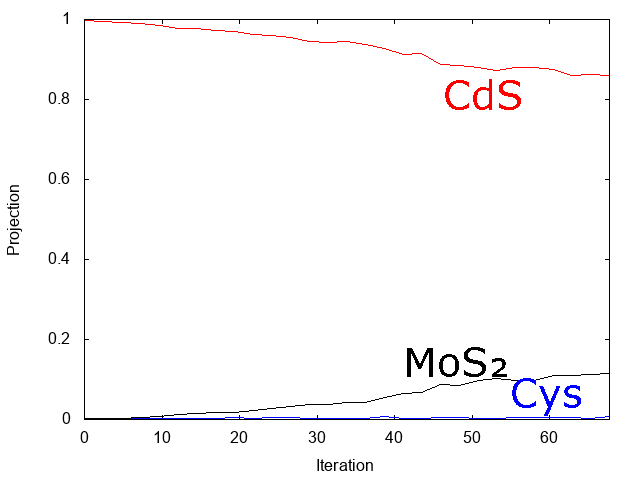}
\vspace{-50pt}\caption{}\vspace{15pt}
\includegraphics[width=\textwidth]{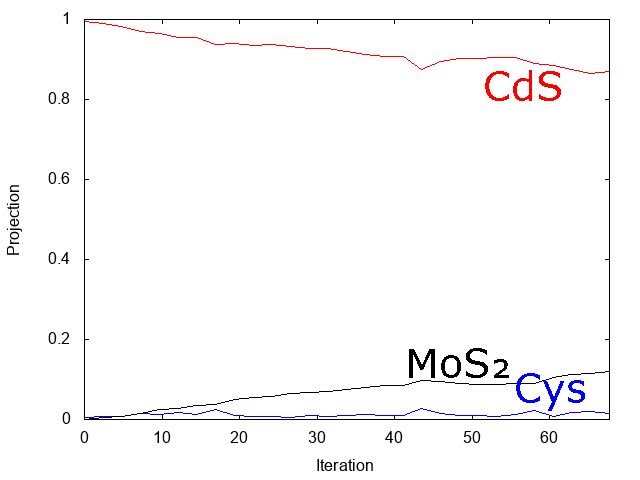}
\vspace{-50pt}\caption{}\vspace{15pt}
\end{subfigure}
\begin{subfigure}[b]{0.306\textwidth}
\includegraphics[width=\textwidth]{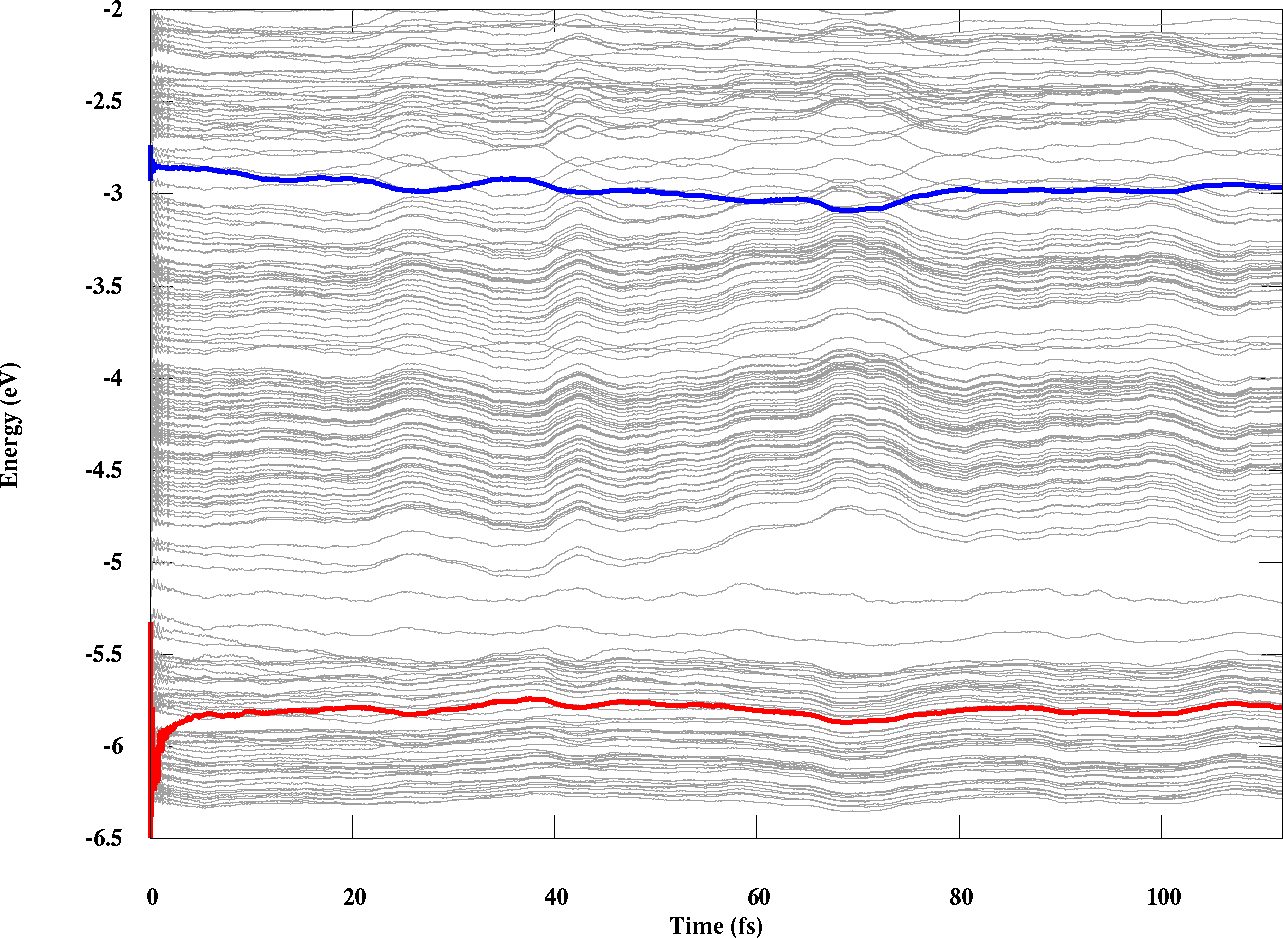}
\vspace{-60pt}\caption{}\vspace{23pt}
\end{subfigure}
\begin{subfigure}[b]{0.162\textwidth}
\includegraphics[width=\textwidth]{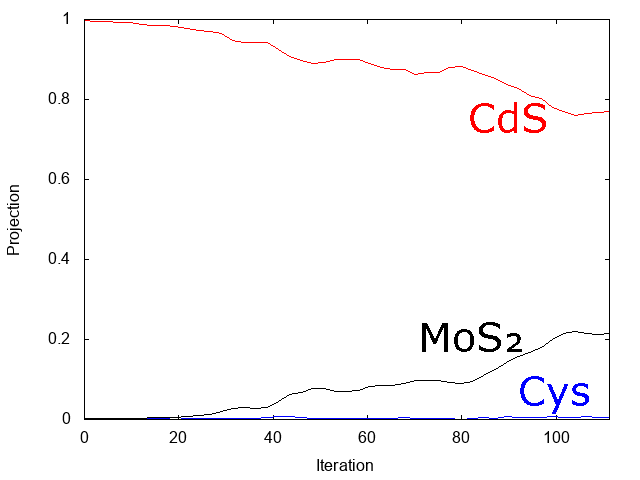}
\vspace{-50pt}\caption{}\vspace{15pt}
\includegraphics[width=\textwidth]{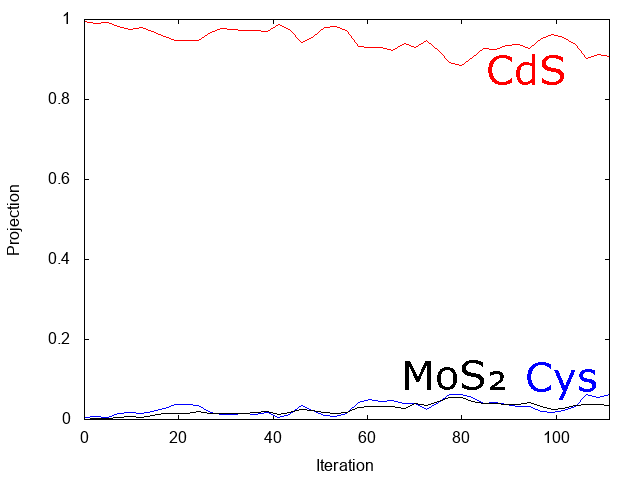}
\vspace{-50pt}\caption{}\vspace{15pt}
\end{subfigure}
\end{center}
\caption{TD-DFT simulation of CdS--Cys($r$)--MoS$_2$-v$_S$ at T=300K. An excitation of $2.00e$ and $0.65e$ from the CdS VBM-2 to CdS CBM+2 was performed in (a-c) and (d-f), respectively. (a,d) Time evolution of the eigenvalues. VBM states below -6 eV are ommitted. The blue line is the excited electron, and the red line is the excited hole. Projections (b,e)/(c,f) of the time-evolved electron/hole state, respectively, onto CdS, Cys, and MoS$_2$.}
\label{fig:tddft1}
\end{figure}

Fig.\ref{fig:tddft1}(a-c) shows the result for the first TD-DFT simulation. The excitation removes two electrons a CdS VB state and promotes it to a CdS CB state. At $t=0$, we see from (a) that the energy level of the hole instantaneously decreases. This is a single-particle picture error, since the eigenfunctions will feel a new charge density. The hole is localized on the S-terminated surface of CdS, where it feels the gradient of the resulting dipole more than the delocalized hole state, which is why it is disproportionately affected at $t=0$. As time evolves, the hole evolves so that its eigenvalue increases, and the excited electron decreases in energy, according to expectations. The collective oscillations in the grey state (CdS VB states and MoS$_2$ CB states) are due to oscillating dipoles created by the movement of the charge density. Notice that when the MoS$_2$ CB band states move down in energy, their CdS VB counterparts move up in energy (and vise versa), since the dipole shifts the relative energy between the two.

Looking at the plots of the projections (b-c), the excited electron state reaches 11.5\% MoS$_2$ character, and the exited hole reaches 12.1\% MoS$_2$ character. If we suppose that the evolution of these states mimics how photoexcited carriers would transfer, then 11.5\% (12.1\%) of the electron (hole) has transferred to MoS$_2$. This contradicts the expectations from band alignment, because based on the band alignment, the hole should not be able to transfer from CdS to MoS$_2$. The reason that it transferred anyway is due to the energy drop at $t=0$. Looking at the density of states in Fig.\ref{fig:tddft-pdos} (a), the MoS$_2$ valence band starts about 1.5 eV lower than the CdS VBM. This means that, at the energy the hole is actually at after $t=0$, there \emph{are} hole states nearby in energy that it can hybridize with. 

For this reason, Fig.\ref{fig:tddft1}(d-f) considers the same excitation but with a magnitude of 0.65. While the energy of the hole does still drop at $t=0$, it does not cross into the MoS$_2$ VB states. The energy levels near the beginning of the simulation also indicate that the amount of energy the hole drops by is directly proportional to the magnitude of the excitation. This is consistent with the explanation that the drop is the hole energy is due to the change in the dipole, since the change in the dipole is proportional to the amount of charge displacement which is proportional to the magnitude of the excitation.

\begin{figure}[H]
\begin{center}
\begin{subfigure}[b]{0.306\textwidth}
\includegraphics[width=\textwidth]{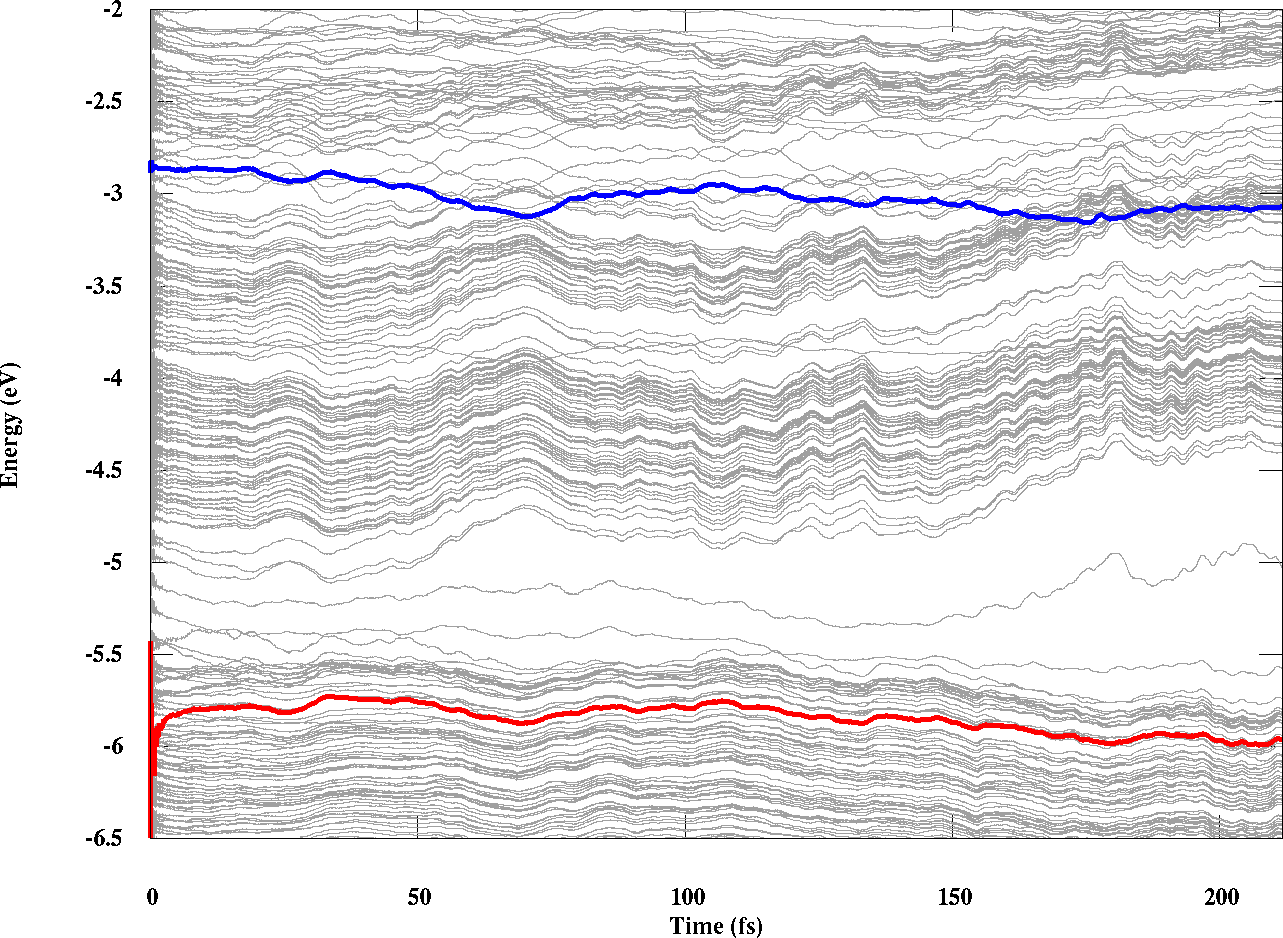}
\vspace{-60pt}\caption{}\vspace{15pt}
\end{subfigure}
\begin{subfigure}[b]{0.162\textwidth}
\includegraphics[width=\textwidth]{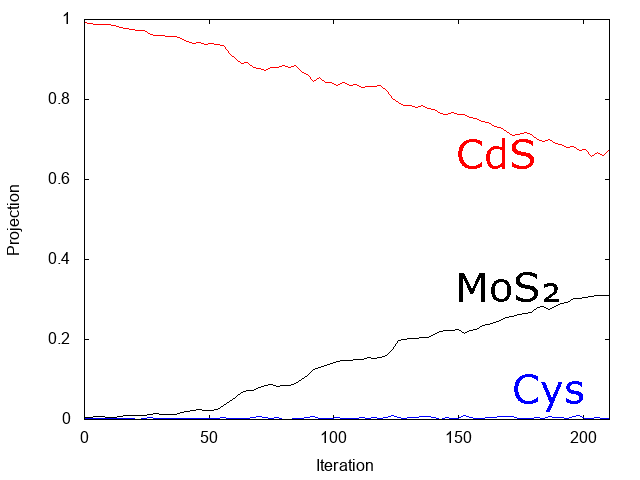}
\vspace{-50pt}\caption{}\vspace{20pt}
\includegraphics[width=\textwidth]{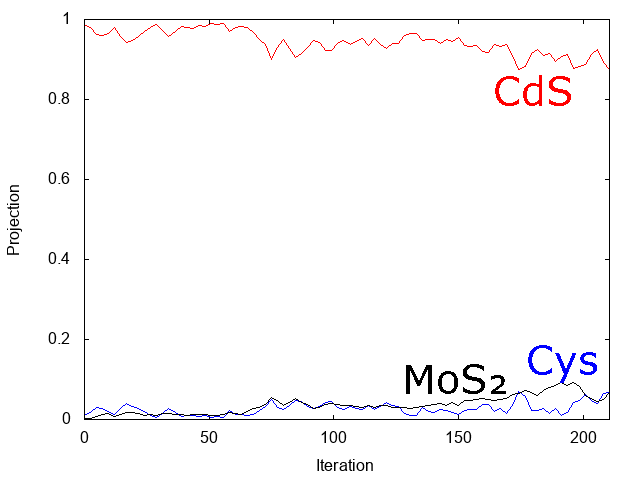}
\vspace{-50pt}\caption{}\vspace{2pt}
\end{subfigure}
\end{center}
\caption{TD-DFT simulation of CdS--Cys($r$)--MoS$_2$-v$_S$ at T=300K. An excitation of $0.65e$ from the CdS VBM-3 to CdS CBM+3 was performed. (a-c) similar to Fig. \ref{fig:tddft1}(a-c).}
\label{fig:tddft2}
\end{figure}

In Fig.\ref{fig:tddft1} (e-f), the electron transfers to MoS$_2$, but the hole does not. This is consistent with the previous observation about the position of the hole relative to the MoS$_2$ VB states. Note that the timescale for Fig.\ref{fig:tddft1}(d-f) is larger than for Fig.\ref{fig:tddft1}(a-c), so the transfer of the electron to MoS$_2$ has reached a larger value of 21.5\%. Fig.\ref{fig:tddft2} shows a similar excitation to the previous one, with two differences: (i) the particular bands used in the excitation are different, but still have CdS projection character, and (ii) the length of the simulation is increased. The overall behavior of the simulation is similar to Fig.\ref{fig:tddft1}(d-f), which gives evidence for the result not being sensitive to the particularly chosen excitation states, but rather their general character. By the end of the simulation, the electron had transferred 31.0\% to MoS$_2$. Interestingly, some fluctuations in the projection of the hole onto Cys can be seen. But from these results, it isn't yet clear how the Cys is playing a role in the electron transfer. The projection plots show that the projection of the excited electron does not reach significant values throughout the simulation.

\begin{figure}[H]
\begin{center}
\begin{subfigure}[b]{0.306\textwidth}
\includegraphics[width=\textwidth]{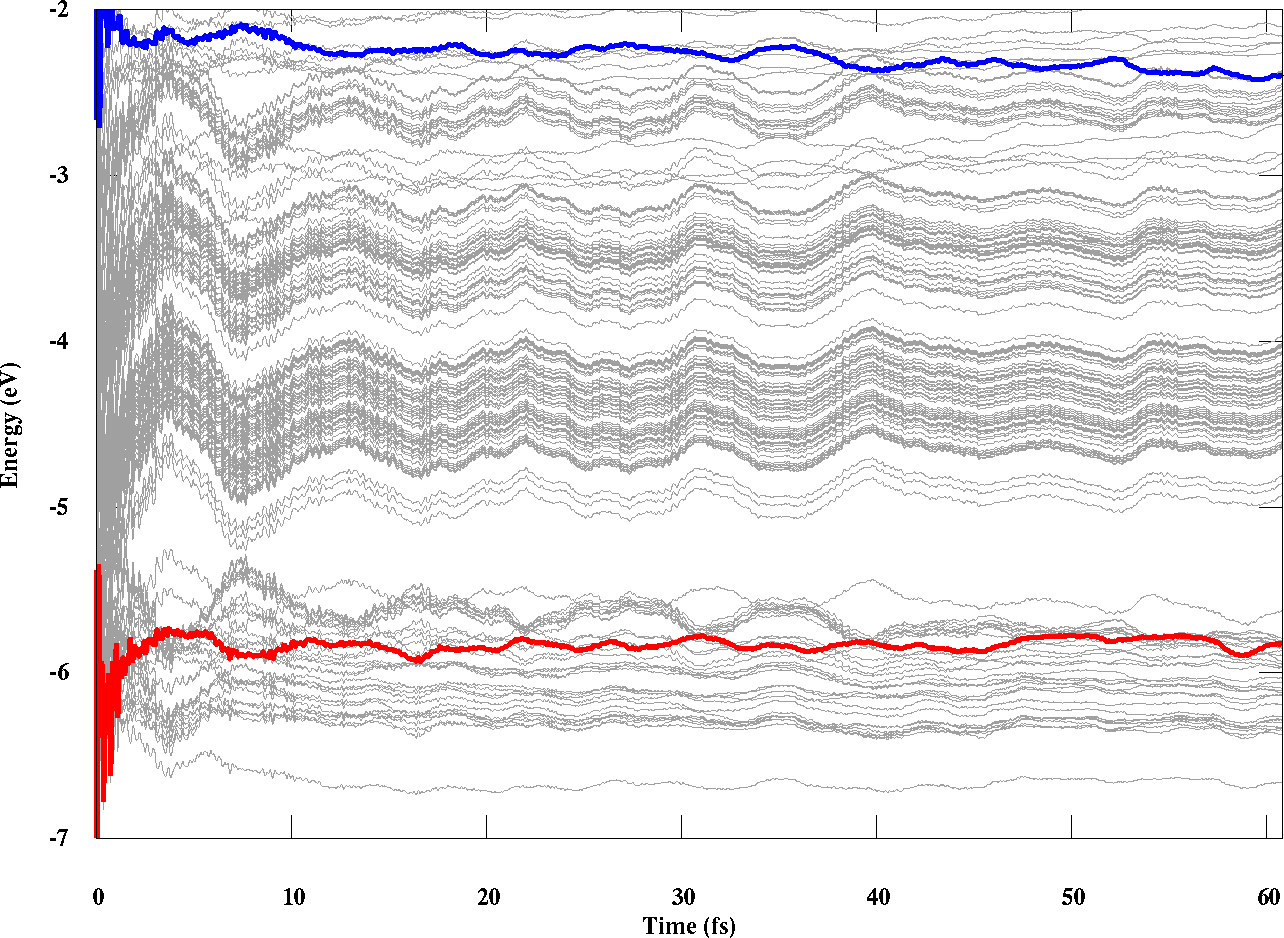}
\vspace{-60pt}\caption{}\vspace{23pt}
\end{subfigure}
\begin{subfigure}[b]{0.162\textwidth}
\includegraphics[width=\textwidth]{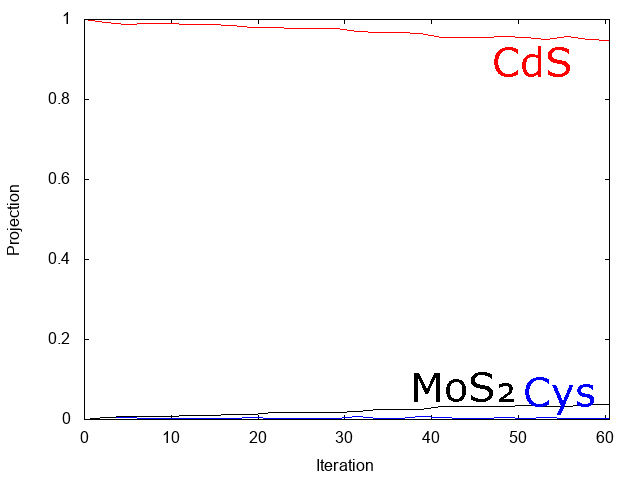}
\vspace{-50pt}\caption{}\vspace{15pt}
\includegraphics[width=\textwidth]{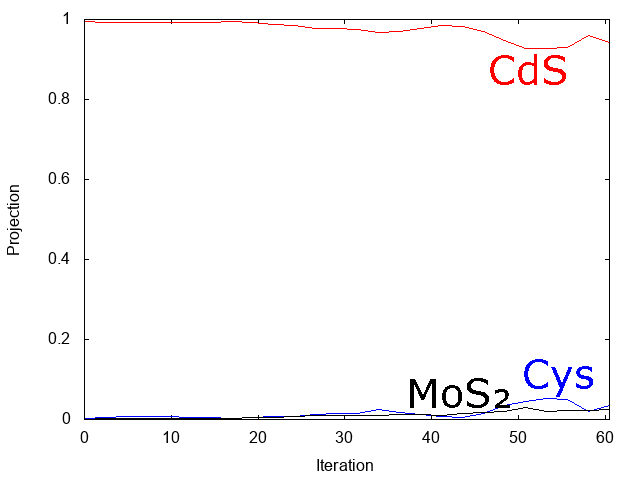}
\vspace{-50pt}\caption{}\vspace{15pt}
\end{subfigure}
\begin{subfigure}[b]{0.306\textwidth}
\includegraphics[width=\textwidth]{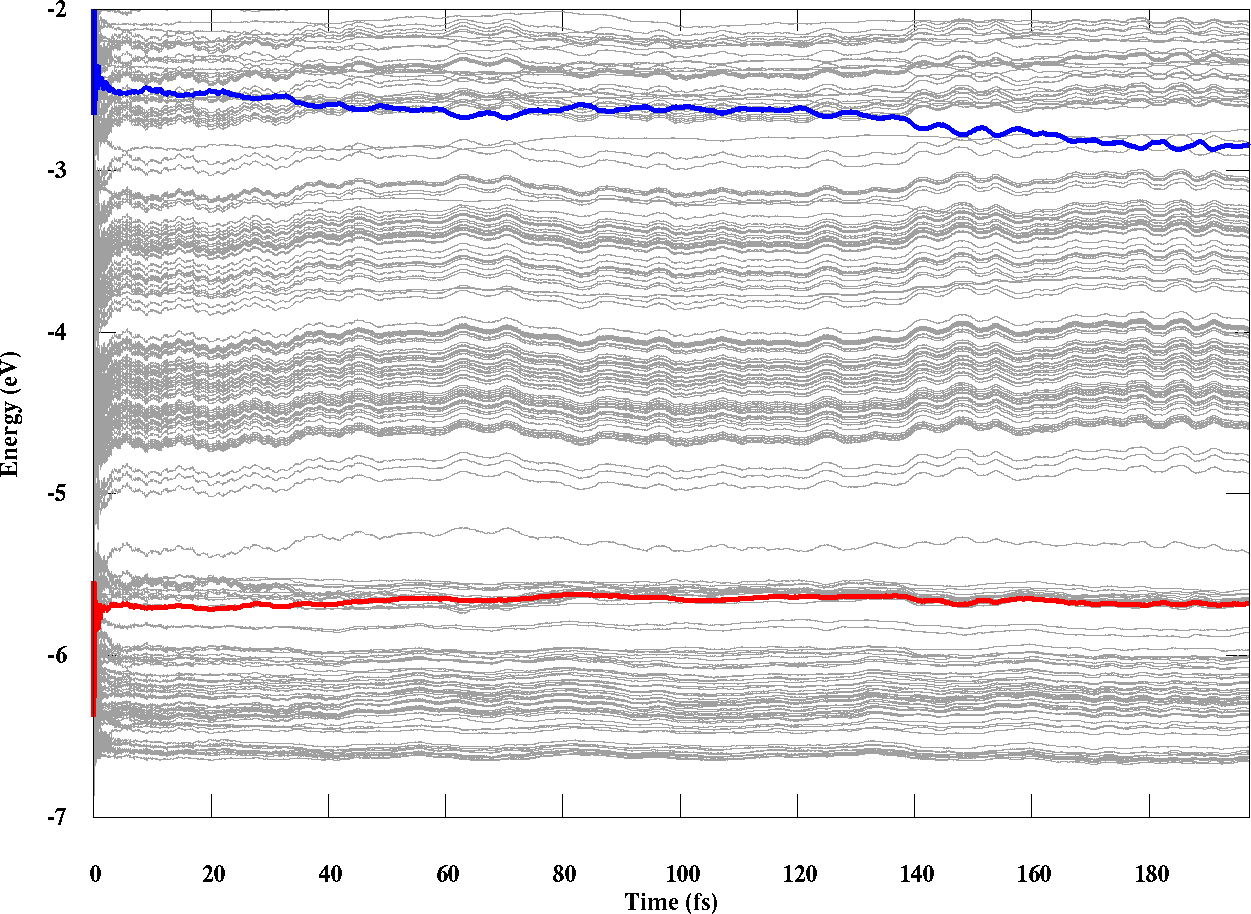}
\vspace{-60pt}\caption{}\vspace{23pt}
\end{subfigure}
\begin{subfigure}[b]{0.162\textwidth}
\includegraphics[width=\textwidth]{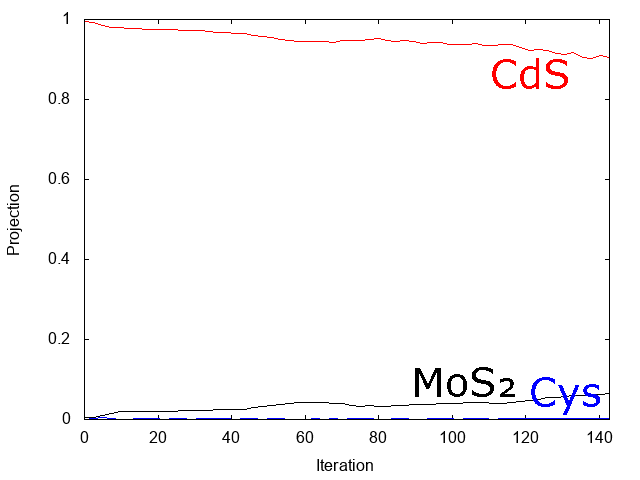}
\vspace{-50pt}\caption{}\vspace{15pt}
\includegraphics[width=\textwidth]{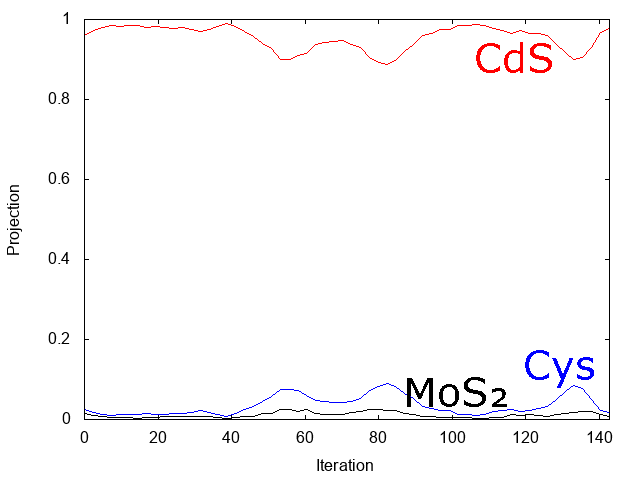}
\vspace{-50pt}\caption{}\vspace{15pt}
\end{subfigure}
\end{center}
\caption{TD-DFT simulation of CdS--Cys($r$)--MoS$_2$-v$_S$ at T=0K. An excitation of $2.00e$ and $0.65e$ from the CdS VBM-1 to CdS CBM+3 was performed in (a-c) and (d-f) respectively. (a,d) and (b-c,e-f) similar to Fig. \ref{fig:tddft1}.}
\label{fig:tddft3}
\end{figure}

\subsection{Results for 0 K Cys Heterostructures}

Fig.\ref{fig:tddft3} shows similar TD-DFT simulations for the T=0K system. Generally, we see that the magnitude of the charge transfer in the T=0K systems is less than the T=300K systems. For example, comparing Fig.\ref{fig:tddft3}(e) with Fig.{fig:tddft2}(b), we see that at a similar total simulation time, the excited electron transferred nearly 40\% in the 300K system but less than 5\% in the 0K system. The differences in charge transfer rate may be due to the out-of-equilibrium atomic positions in the 300K system causing broadening in the density of states and subsequent coupling between these states. Another noticeable observation is that the magnitude of the hole energy drop is less in the 0K system than in the 300K system. It isn't clear why that should be the case, but similar observations have occured in TD-DFT simulations of other systems \cite{cheng2020carrier}. Also, the lack of coupling causes coherent waves in the fluctuations of charge density, as seen from the relatively steady-frequency oscillations in Fig.\ref{fig:tddft3}(a) and (d).

\begin{figure}[H]
\begin{center}
\begin{subfigure}[b]{0.306\textwidth}
\includegraphics[width=\textwidth]{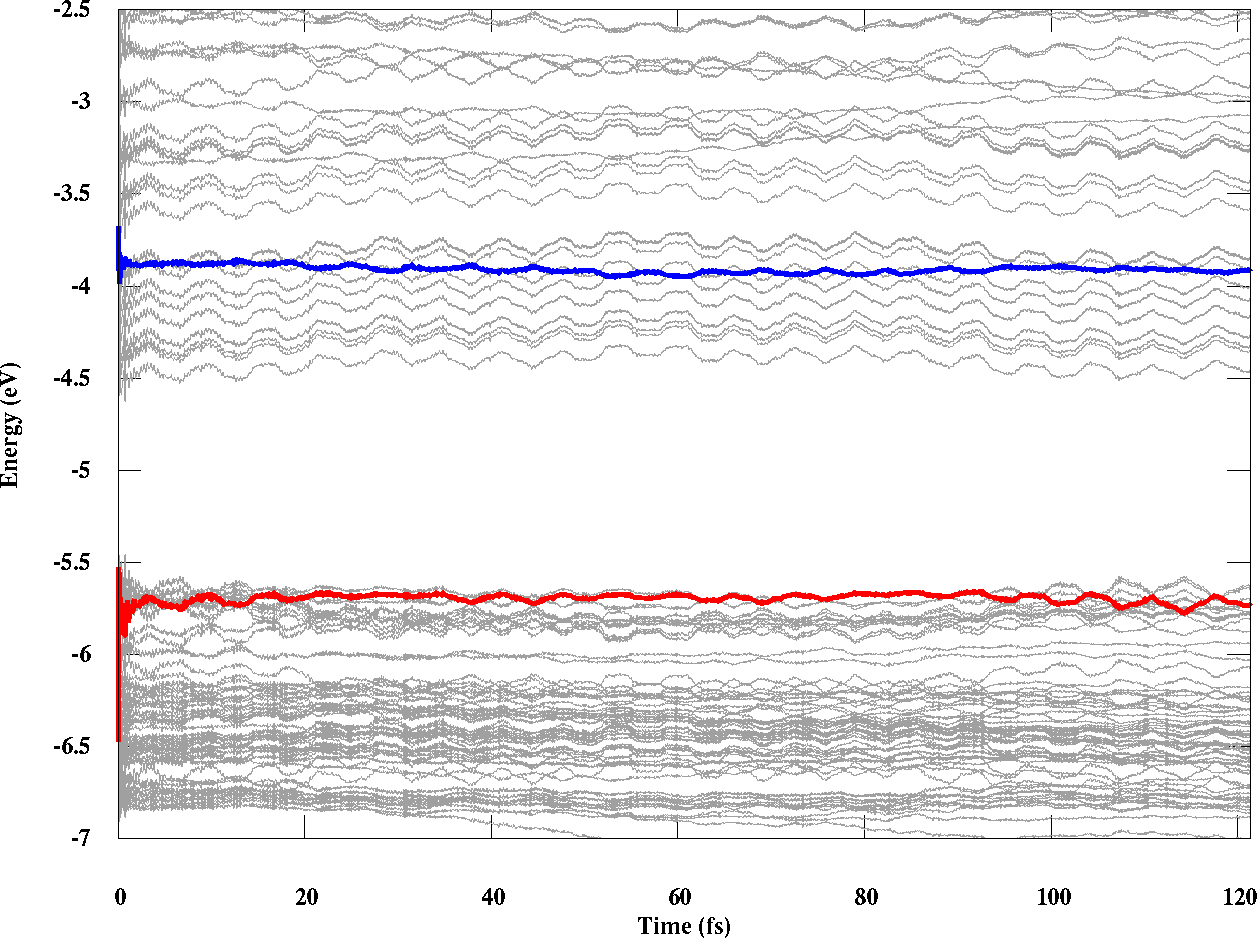}
\vspace{-60pt}\caption{}\vspace{23pt}
\end{subfigure}
\begin{subfigure}[b]{0.162\textwidth}
\includegraphics[width=\textwidth]{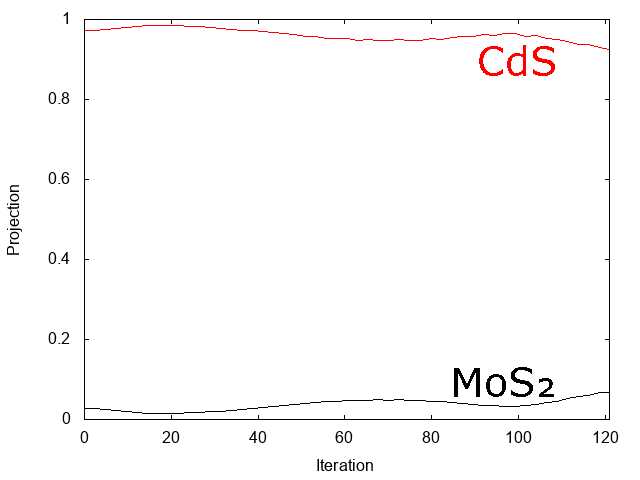}
\vspace{-50pt}\caption{}\vspace{15pt}
\includegraphics[width=\textwidth]{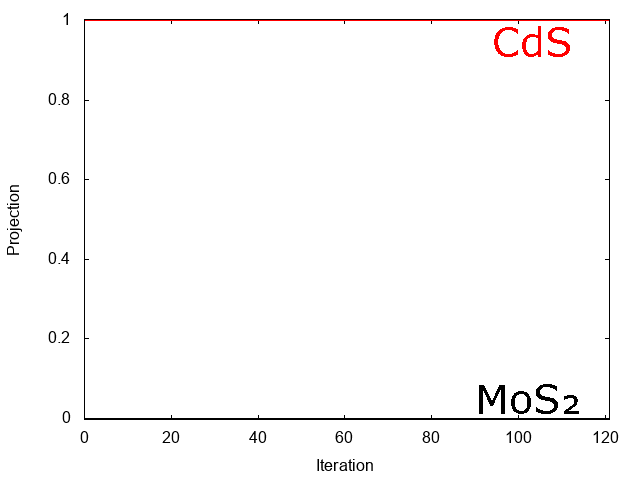}
\vspace{-50pt}\caption{}\vspace{15pt}
\end{subfigure}
\begin{subfigure}[b]{0.306\textwidth}
\includegraphics[width=\textwidth]{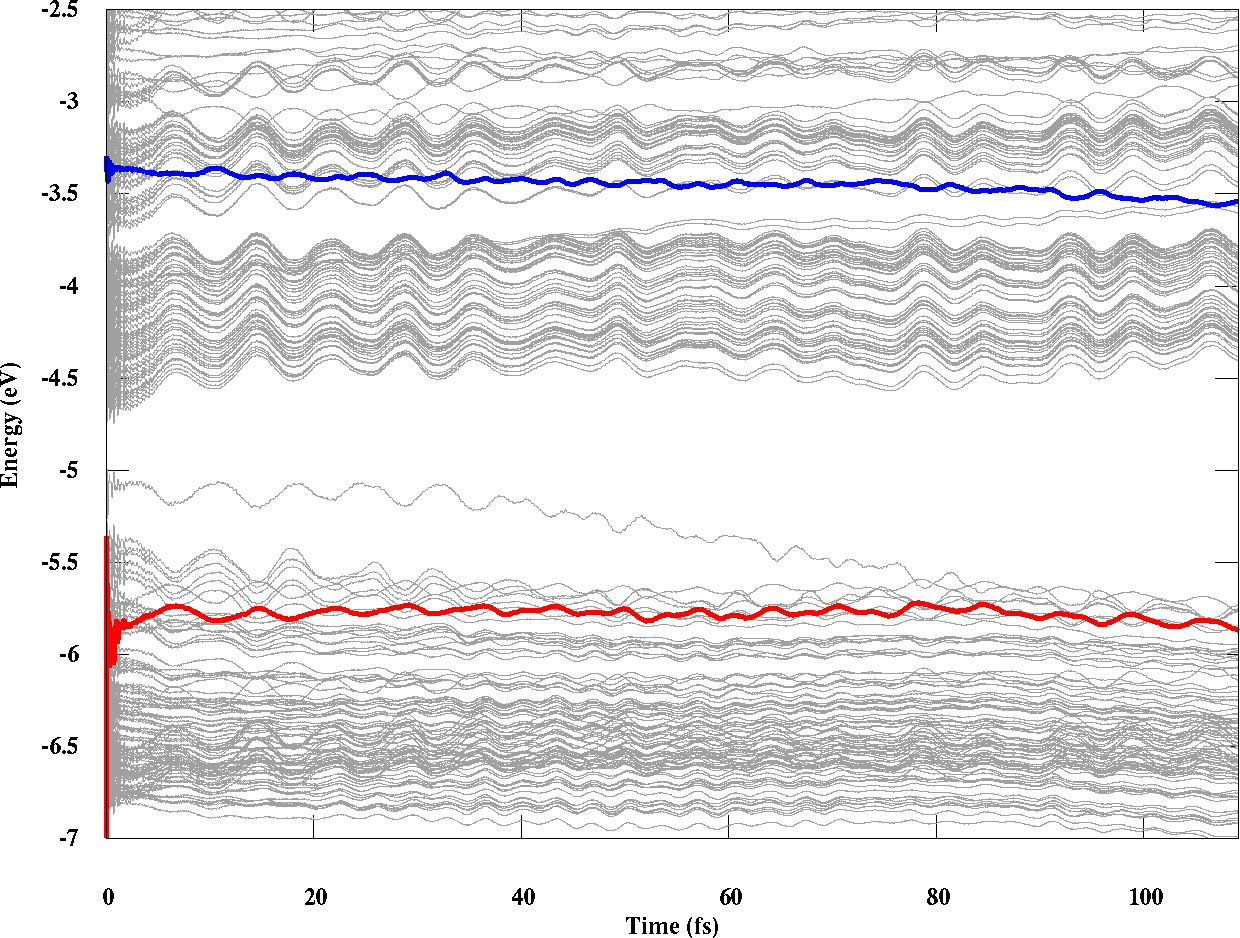}
\vspace{-60pt}\caption{}\vspace{23pt}
\end{subfigure}
\begin{subfigure}[b]{0.162\textwidth}
\includegraphics[width=\textwidth]{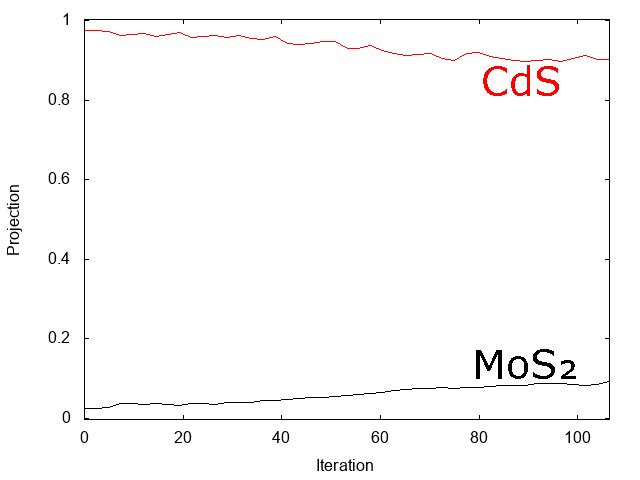}
\vspace{-50pt}\caption{}\vspace{15pt}
\includegraphics[width=\textwidth]{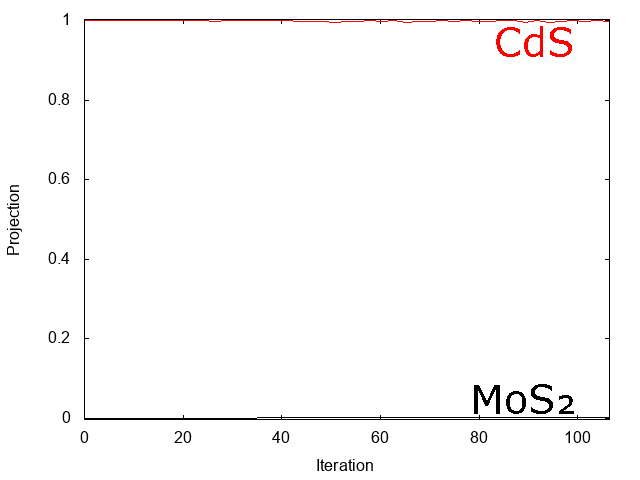}
\vspace{-50pt}\caption{}\vspace{15pt}
\end{subfigure}
\end{center}
\caption{TD-DFT simulation of CdS--MoS$_2$ at T=0K and T=300K for (a-c) and (d-f) respectively. An excitation of $0.65e$ from the CdS VBM-1 (VBM-1) to CdS CBM (CBM+1) was performed in (a-c) and (d-f) respectively. (a,d) and (b-c,e-f) similar to Fig. \ref{fig:tddft1}.}
\label{fig:tddft4}
\end{figure}

\subsection{Results for non-Cys Heterostructures}

Next, we consider the simulation of systems that do not have a ligand molecule in the middle. The distance between the CdS and MoS$_2$ films was fixed to be same as the ligated heterostructures, namely 5.504 {\AA} for the 0K system and 5.329 {\AA} for the 300K system. In Fig.\ref{fig:tddft4}, it is clear from the projection of the time-evolved electron (b,e) and hole (c,f) states that there is a significant reduction in the electron transfer rate. In fact, an imperceptible amount of the excited hole has transferred to MoS$_2$ in these systems, in contrast with the modest fluctuations seen in the ligated heterostructures. Comparing the electron transfer rates between the two systems, we see that the ligand enhances the electron rate by approximately a factor of 2.2. This enhancement may be underestimated due to the non-negligible MoS$_2$ projection of the initial acceptor state in the non-ligand TD-DFT simulations.

\subsection{Conclusion}

We find that TD-DFT successfully predicts the charge separation process seen in CdS heterostructures, namely that the excited electron in CdS transfers to MoS$_2$. This transfer rate is greater when the system is at an elevated ionic temperature, and is sensitive to the dipole resulting from the photoexcited charge density and thus the cross-sectional excitation density. The charge transfer rates that are observed in heterostructures with the ligand molecule are greater than in the non-ligated counterparts by a factor of about 2.2.

While these TD-DFT simulations have successfully shown that the presence of the ligand molecule improves the charge transfer, little time-evolved projection onto Cys was observed. One explanation is that the ligand density of states, which is hybridized with CdS and MoS$_2$ over a wide range of energies, effectively couples these states together, and this coupling has significant effects even if the magnitude of the coupling is small.


\end{multicols}

\bibliography{TDDFT_CdS_heterostructures}
\bibliographystyle{ieeetr}

\end{document}